\pdfoutput=1

\documentclass[11pt]{article}

\usepackage[final]{acl}

\usepackage{times}
\usepackage{graphicx}
\usepackage{latexsym}
\usepackage{booktabs}

\usepackage[T1]{fontenc}

\usepackage[utf8]{inputenc}

\usepackage{microtype}
\usepackage{hyperref}
\usepackage {subcaption}

\usepackage{inconsolata}

\usepackage{graphicx}
\usepackage{amssymb}
\usepackage{amsmath}
 \usepackage{multirow}

\title{A Generative Adaptive Replay Continual Learning Model for \\ Temporal Knowledge Graph Reasoning}

\author{
 \textbf{Zhiyu Zhang\textsuperscript{1,3}},
 \textbf{Wei Chen\textsuperscript{2\thanks{Corresponding author}}},
 \textbf{Youfang Lin\textsuperscript{1,3}},
 \textbf{Huaiyu Wan\textsuperscript{1,3}}
\\
 \textsuperscript{1}School of Computer and Information Technology, Beijing Jiaotong University, Beijing, China
 \\
 \textsuperscript{2}Guilin University of Electronic Technology, 
 \\ School of Computer Science and Information Security, Guangxi, China
 \\
 \textsuperscript{3}Beijing Key Laboratory of Traffic Data Analysis and Mining, Beijing, China
\\
 \texttt{\{zyuzhang, yflin, hywan\}@bjtu.edu.cn}, 
 \texttt{w\_chen@guet.edu.cn}
 \\
}

\begin{document}
\maketitle
\begin{abstract}
Recent Continual Learning (CL)-based Temporal Knowledge Graph Reasoning (TKGR) methods focus on significantly reducing computational cost and mitigating catastrophic forgetting caused by fine-tuning models with new data. However, existing CL-based TKGR methods still face two key limitations: (1) They usually one-sidedly reorganize individual historical facts, while overlooking the historical context essential for accurately understanding the historical semantics of these facts; (2) They preserve historical knowledge by simply replaying historical facts, while ignoring the potential conflicts between historical and emerging facts.  In this paper, we propose a \textbf{D}eep \textbf{G}enerative \textbf{A}daptive \textbf{R}eplay (DGAR) method, which can generate and adaptively replay historical entity distribution representations from the whole historical context. To address the first challenge, historical context prompts as sampling units are built to preserve the whole historical context information. To overcome the second challenge, a pre-trained diffusion model is adopted to generate the historical distribution. During the generation process, the common features between the historical and current distributions are enhanced under the guidance of the TKGR model. In addition, a layer-by-layer adaptive replay mechanism is designed to effectively integrate historical and current distributions. Experimental results demonstrate that DGAR significantly  
outperforms baselines in reasoning and mitigating forgetting.

\end{abstract}

\section{Introduction}

Temporal Knowledge Graphs (TKGs) extend traditional Knowledge Graphs (KGs) by associating triples with timestamps~\citep{TTransE,HyTE,conf/iclr/LacroixOU20}, providing dynamic and structured time-sensitive knowledge for various downstream applications ~\citep{LLM_TKG, gutierrez2024hipporag, wang2024large, Zhao0SZLW25}, such as Large Language Models reasoning, event prediction, and financial forecasting~\citep{guan2022event}.
Unfortunately, TKGs often suffer from incompleteness, hindering the capability of dynamic knowledge representation in downstream applications. Temporal knowledge Graph Reasoning (TKGR) is proposed to address this issue by inferring missing temporal facts based on historical knowledge.

In real-world scenarios, TKGs are continuously updated with unseen entities, relations, and new facts. 
Existing TKGR studies~\citep{TTransE,RE-GCN,TiRGN,CENET} update model parameters by retraining on the entire TKG when new data arrives.
This process is computationally expensive and impractical for dynamic settings, especially in the transportation and finance domains where frequent knowledge updates are required ~\citep{IncDE}.
Continual Learning(CL) fine-tuning models with new data may seem intuitive, but often results in catastrophic forgetting, where prior knowledge is lost~\citep{HistoryRepeat}. 

To mitigate catastrophic forgetting, recent CL-based TKGR studies ~\citep{wu2021tie,HistoryRepeat}  rely on the mechanisms of replaying prior knowledge, further employing regularization techniques to preserve old knowledge. 
These studies integrate new knowledge while preserving previously acquired information, thus enabling reasoning over both historical and emerging data.

Despite notable progress, the currently  CL-based TKGR methods still face two primary challenges: 

(1) These methods often reorganize and replay the historical data (e.g., based on frequency or clustering) to mitigate catastrophic forgetting.
However, such methods solely focus on the statistical properties of individual historical events, which fails to correctly understand the historical semantics of these facts combining the necessary historical context. Besides, this fragmented approach makes it difficult to capture the overarching trends of entity behavior, thus limiting the model's capacity in complex reasoning tasks.

\begin{figure}[t]
    \centering
    \includegraphics[width=\columnwidth,height=0.17\textheight]{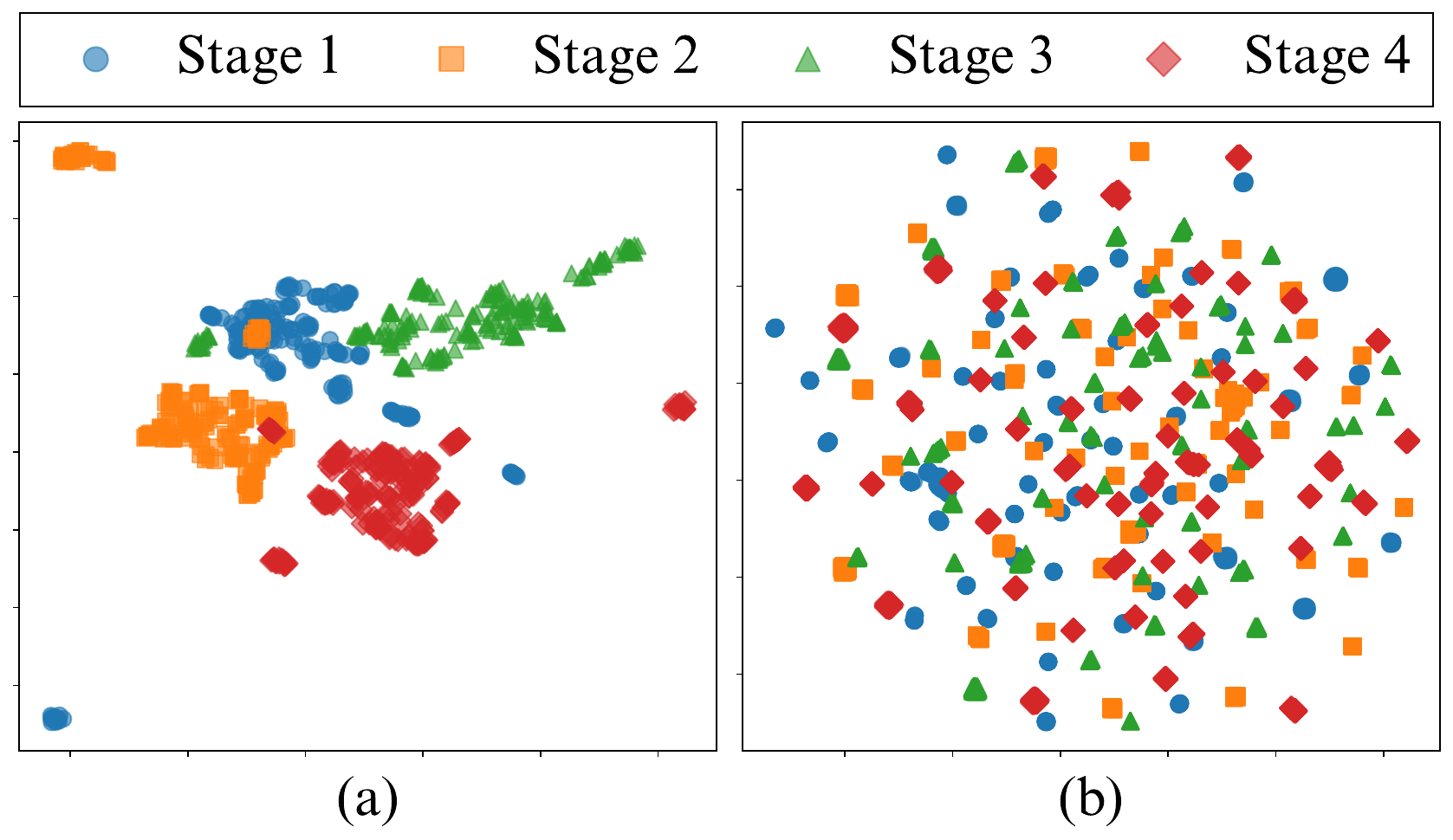}
    \caption{The distributions of the same set of entity features at different timestamps are visualized using U-MAP visualization. Entities involved in all facts at a randomly selected time $i$ are extracted. Stage 1 represents the feature distribution of these entities at time $i$, while stage 2, 3, and 4 respectively correspond to their feature distributions at times $j$, $m$ and $n$, where $i < j < m < n$. (a) depicts the distributions learned by the base model across these timestamps, and (b) shows the distributions learned by DGAR. The results demonstrate that our approach effectively resolves distribution conflicts while preserving historical knowledge.}  
    \label{fig:mainfig}
    \vspace{-1.7em}
\end{figure}

(2) Current approaches typically replay historical data directly overlooking potential conflicts between the distributions of historical and current data (e.g., Figure \ref{fig:mainfig}). As entities associate with different neighbors over time, semantic differences arise, which in turn cause conflicts in the distributions of entities at different times. This oversight hinders the effectiveness of mitigating catastrophic forgetting.
    
To address these challenges, we propose a Deep Generative Adaptive Replay (DGAR) method for TKGR, which can continually and adaptively replay historical information by generating the historical distribution representation of entities from the whole historical context. 
For the first challenge, instead of using individual facts as sampling units, we build Historical Context Prompts (HCPs) as sampling units to retain the context information of historical data. 
For the second challenge, we enhance the common features across different distributions and introduce a deep adaptive replay mechanism to mitigate distribution conflicts. Specifically,
we design a Diffusion-Enhanced Historical Distribution Generation (Diff-HDG) strategy that generates entity historical distribution representations. During the generation process, the features of the entity's historical distribution that are common to the entity's current distribution are enhanced.
In addition, a layer-by-layer Deep Adaptive Replay (DAR) mechanism is introduced to inject the entity's historical distribution representation into its current distribution representation.
\footnotetext[1]{The source code of DGAR is available at: \href{https://github.com/zyzhang11/DGAR}{https://github.com/zyzhang11/DGAR}.}
In summary, the main contributions of this work are as follows:
\vspace{-0.5em}
\begin{itemize}
\item[$\bullet$] We propose a novel Generative Adaptive Replay Continual Learning method for TKGR, which effectively addresses the issue of knowledge forgetting by incorporating the entire historical context and mitigating distribution conflicts.
\vspace{-0.7em}
\item[$\bullet$] A sophisticated historical context prompt is designed for replay data sampling, ensuring the semantic integrity of the historical context information in the sampled facts.
\vspace{-0.5em}
\item[$\bullet$]A Diff-HDG strategy is proposed to generate historical distribution representations by enhancing the common features. In addition, a DAR mechanism is designed to efficiently integrate historical and current distributions.
\vspace{-0.5em}
\item[$\bullet$]Extensive experiments conducted on widely used TKGR datasets demonstrate the superiority of our approach, consistently outperforming all baseline methods across various metrics.
\end{itemize}

\vspace{-0.9em}
\section{Related Work}
\vspace{-0.2em}
\subsection{Reasoning on TKGs}
 TKGR aims to infer missing facts by utilizing known facts. Recent advancements in this field fall into four main approaches. The distribution-based approaches~\citep{TTransE,conf/iclr/LacroixOU20} perform reasoning by training a scoring function that can evaluate the distance or semantic similarity between entities.
 The Graph Neural Network (GNN)-based methods~\citep{RE-GCN, conf/ijcai/LiS022,xu-etal-2023-cenet, Log_CL, towards-enhancing, Chen2024CognTKEAC} capture structural and temporal patterns in graph sequence to enhance reasoning accuracy. 
 Rule-based temporal knowledge graph reasoning methods~\citep{huang-etal-2024-confidence,chen-etal-2024-unified} follow a symbolic paradigm that emphasizes interpretability, logical consistency, and low resource requirements. These methods typically mine temporal logical rules from historical fact sequences, then use these rules to infer future events or fill in missing historical facts. 
When new data arrives, these methods often require retraining.  Given the strong performance of GNN-based methods in TKGR, our approach builds upon this category of methods.

\vspace{-0.5em}
\subsection{CL for Knowledge  Graphs}
\vspace{-0.3em}
Compared to existing approaches that necessitate repeated retraining, CL adaptively incorporates sequentially evolving knowledge.
Recently, several methods have applied CL to knowledge graph embedding (KGE) and TKGR. For instance, some approaches~\citep{wu2021tie,HistoryRepeat} integrate experience replay with regularization techniques to address catastrophic forgetting in TKGR. TIE's~\citep{wu2021tie} overly restrictive regularization leads to a decline in overall performance. The regularization method restricts the applicability of DEWC~\citep{HistoryRepeat} to a limited number of tasks. ~\citep{cui2023lifelong,IncDE,liu2024fast} apply CL to KGE by employing regularization constraints to retain historical knowledge, effectively mitigating catastrophic forgetting.
\vspace{-0.3em}
\subsection{Diffusion Models}
\vspace{-0.2em}
Diffusion models are generative frameworks that reconstruct structured data from Gaussian noise through a stepwise reverse denoising process~\citep{sohl2015deep,ho2020denoising}. \textbf{In continuous domains} like image synthesis, DDPM and its variants effectively model complex distributions and generate high-quality outputs~\citep{ho2020denoising,rombach2022high}. Applying diffusion models to \textbf{discrete domains} is challenging due to Gaussian noise's incompatibility with discrete structures. Text generation employs polynomial diffusion or continuous-to-discrete mapping to link continuous processes with discrete data~\citep{austin2021structured,gong2022diffuseq,li2022diffusion}. \textbf{In KGs}, ~\citep{long2024fact,cai2024predicting} restore knowledge from noise by mapping discrete KG data to a continuous space and applying conditional constraints.
\vspace{-1.8em}
\section{Preliminaries}
\vspace{-0.4em}
\subsection{The Task of TKGR}
TKG  can be represented as a sequence of snapshots partitioned by time, denoted as $\mathcal{G}=\left\{{G}_1,{G}_2,{G}_3,...,{G}_T\right\}$. 
Each snapshot ${G}_{t}=(\mathcal{V},\mathcal{R},\mathcal{F}_{t})$ is a directed multi-relational graph at timestamp $t$.
$(s,r,o,t) \in$ $\mathcal{F}_{t}$ is denoted as a fact,
where  $s\in \mathcal{V}$ and $o \in \mathcal{V}$ are subject entity and object entity, respectively, $r \in \mathcal{R}$ is denoted as a relation that connects the subject entity and the object entity.

The task of TKGR aims to predict the missing object entity (or subject entity) given a query $({s}_q,{r}_q,?,{t}_q)$ or $(?,{r}_q,{o}_q,{t}_q)$. To be consistent with common representation, the inverse quadruple of a fact $(s,r,o,t)$ is $(o,{r}^{-1},s,t)$ which is added to the dataset. 
The TKG reasoning goal can be expressed as the prediction of object entities.

\subsection{Continual Learning for TKGR}
Under the CL setting, TKGs can be viewed as a sequence of KG snapshots arriving as a stream over time. A set of tasks can be denoted as $\left\{\mathcal{T}_{1},\mathcal{T}_{2},...,\mathcal{T}_{T}\right\}$, where each task is denoted as $\mathcal{T}_{t}=(D_{train}^{t}, D_{valid}^{t}, D_{test}^{t})$, where $G_{t}=[D_{train}^{t}: D_{valid}^{t}: D_{test}^{t}]$. The model parameters are updated sequentially for each task as the task stream $\left\{\mathcal{T}_{1},\mathcal{T}_{2},...,\mathcal{T}_{T}\right\}$ arrives.
The trained model parameters at each step can be represented as $\left\{\theta_{1},\theta_{2},...,\theta_{T}\right\}$. 
At time $t$, the parameters $\theta_t$ are initialized by the parameters $\theta_{t-1}$ at the previous time. Then the model is trained on $D_{train}^t$ to update the parameters.

During CL for TKGR,  we mitigate catastrophic forgetting based on KG snapshot sequence reasoning models, such as RE-GCN~\citep{RE-GCN}.
We focus primarily on entity representations, as the semantics of entities tend to evolve more frequently over time, in contrast to the relatively negligible changes in the semantics of relations~\citep{goel2020diachronic}.



\vspace{-0.3em}
\subsection{Denoising Diffusion Probabilistic Model}
\begin{figure*}
    \centering
    \includegraphics[width=\textwidth]{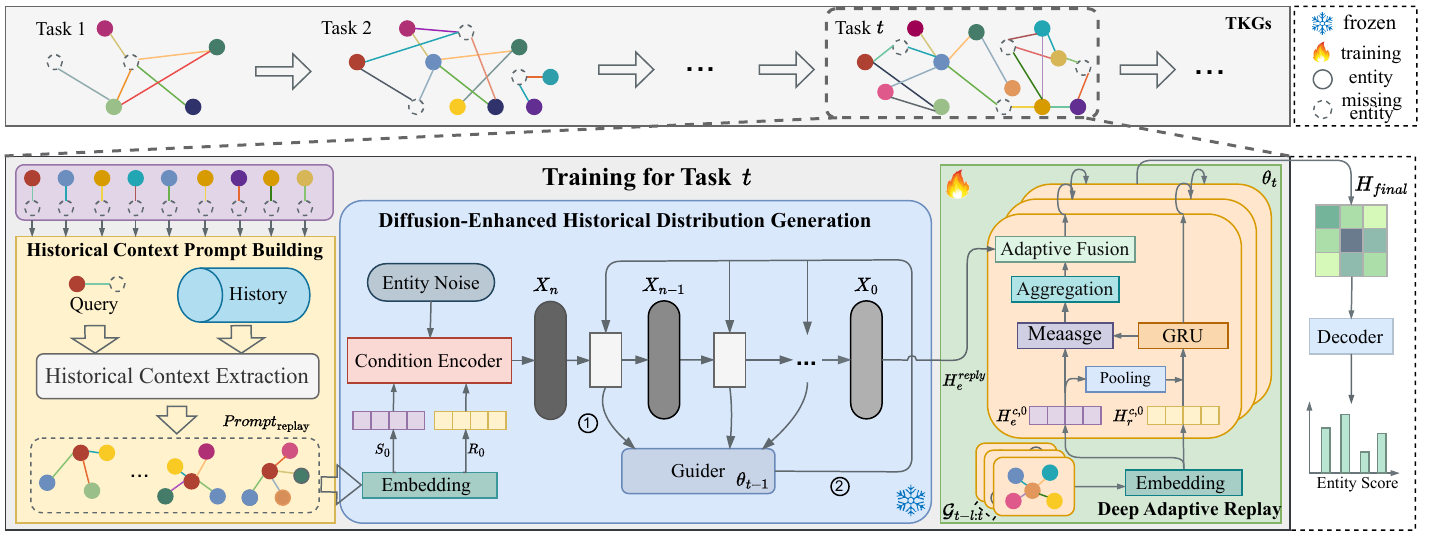}
    \caption{The overall architecture diagram of DGAR. Following the CL paradigm, each snapshot of the TKGs is treated as a separate task.}
    \label{fig:model_image}
    \vspace{-1.5em}
\end{figure*}
The Diffusion Model (DM) consists of a forward diffusion process and a reverse diffusion process. 
In the forward process, a continuous DM is adapted to handle discrete facts $\mathcal{G}$. Given discrete data $x$, we first project $x$ into a continuous embedding, denoted as ${X_0} = \operatorname{Embedding}(x), 
$ where $X_0\in{\mathbb{R}^d}$. $\operatorname{Embedding}(\cdot)$ is a function that can map a word to a vector in $\mathbb{R}^d$. Then a Markov chain of latent variables $X_1,X_2,...,X_n$ is generated in the forward process by gradually adding small amounts of standard Gaussian noise to the sample. This process can be obtained by: 
\begin{equation}
 \vspace{-0.7em}
    \resizebox{0.83\linewidth}{!}{
    \begin{math}
      \begin{aligned}
      q\left(X_{n} \! \mid \!X_{n-1}\right)=\mathcal{N}\left(X_{n} ; \sqrt{1-\beta_{n}} X_{n-1}, \beta_{n} I\right),
      \end{aligned}
    \end{math}
    }
  \label{eq:1}
\end{equation}
where $\beta_{n}, n\in[1,...,N]$ is a noise schedule used to control the step size of the added noise and $I$ is an identity matrix. $\mathcal{N}$ is the Gaussian distribution.

In the reverse process, the standard Gaussian representation $X_t$ progressively approximates the true representation $X_0$ by iterative denoising. It can be learned by a parameterized model:
\begin{equation}
\vspace{-0.1em}
    \resizebox{0.99\linewidth}{!}{
    \begin{math}
      \begin{aligned}
      p_{\phi}\left(\mathrm{X}_{n-1} \! \mid \! X_{n}, n\right)=\mathcal{N}\left(X_{n-1} ; \mu_{\phi}\left(\mathrm{X}_{n}, n\right), \Sigma_{\phi}\left(\mathrm{X}_{n}, n\right)\right),
      \end{aligned}
    \end{math}
    }
  \label{eq:2}
\end{equation}
where $\mu_{\phi}$ and $\Sigma_{\phi}$ are generally implemented by a deep  neural ${f}_\phi(\cdot)$, such as Transformer or U-Net. Inspired by the success of Transformer encoders in the field of graph data\citep{hu2020heterogeneous}, we opt for the Transformer architecture in this work. The pretraining objectives are defined as follows: 
\begin{equation}
    \resizebox{0.99\linewidth}{!}{
    \begin{math}
      \begin{aligned}
      \begin{array}{l}
\mathcal{L}=\mathbb{E}_{q}\left[\sum_{n=2}^{N}\left\|X_{0}-f_{\phi}\left(X_{n}, n\right)\right\|^{2}\right]-\log p_{\phi}\left(x \mid X_{0}\right)
        \end{array},
      \end{aligned}
    \end{math}
    }
  \label{eq:3}
\end{equation}
where $\mathbb{E}_{q}$ is the expectation over joint distribution. It is important to emphasize that the data employed in the testing phase remains entirely unseen during the pretraining process.

\section{The DGAR Method}
\label{sec:DGAR}
The overall architecture of DGAR is shown in Figure \ref{fig:model_image}, primarily consisting of three parts:  Historical Context Prompt Building, Diffusion-Enhanced Historical Distribution Generation, and Deep Adaptive Replay. 

Initially, when a new query arrives for the $t$-th task, DGAR builds HCPs based on the queried entity (Section \ref{sec:HP}). Based on the obtained HCPs, we adopt the latest TKGR model parameters $\theta_{t-1}$ to guide the historical distribution representation generation of entities (Section \ref{sec:FG}). To support the following reasoning, DAR injects generated historical entity distribution into the current entity distribution representation (Section \ref{sec:DFR}). Finally, we present the final loss function of DGAR (Section \ref{sec:Optimization}).

\subsection{Historical Context Prompt Building}
\label{sec:HP}
Historical context prompts, serving as sampling units of replay data, aim
to accurately preserve entities' complete historical semantics.
Before constructing an HCP, it is critical to determine which entities at time $t$ are most relevant to achieving the ultimate goal of mitigating catastrophic forgetting.
As a new query $({e}_q,{r}_q,?,t)$ arrives, ${e}_q$ is an involved entity and its semantics will be directly influenced after fine-tuning at the current timestamp $t$~\citep{StreamE}.

To correctly reflect the historical context semantics of $e_q$, we construct an HCP for entity $e_q$.
If ${e}_q$ appears at time $t-1$ or earlier, it might be associated with one or more entities in the past. The historical distribution of ${e}_q$ is determined by the entities historically associated with ${e}_q$~\citep{xing2024less}. The HCP building for $e_q$ can be formalized as follows: 
\begin{equation}
    \vspace{-0.3em}
    \resizebox{0.75\linewidth}{!}{
    \begin{math}
      \begin{aligned}
      {Prompt}_{\text {replay }}^{i}=\left\{\left(s, r, e_{q}\right) \mid\left(s, r, e_{q}\right) \in G_{i} \right. \\  \left. \text{or} \left(e_{q}, r^{-1}, s\right) \in G_{i} , G_{i} \in \mathcal{G} \right\},
      \end{aligned}
    \end{math}
    }
  \label{eq:1}
\end{equation}
where ${Prompt}_{\text {replay }}^{i}$ is the HCP of entity $e_q$ at time $i$, which denotes the set of triples associated with ${e}_q$ at historical moment $i$. The triples in ${Prompt}_{\text {replay }}^{i}$ consist of the entity $e_q$, the neighbor $s$ associated with $e_q$ at time $i$, and the relation $r$ between $e_q$ and $s$ at time $i$. When no triple containing $e_q$ appears at time $i$, ${Prompt}_{\text {replay }}^{i}$ is empty. 

To reduce the computational and storage burden, we do not select the HCP of $e_q$ across its entire history. Instead, we treat a HCP as the sampling unit, and randomly select HCPs of $k$ distinct time to enhance the generalizability of the replay data.
The discussion about $k$ is provided in Appendix \ref{sec: hype}. The set of HCP after sampling is denoted as ${Prompt}_{\text {replay }}$, which serves as the prompt for generating entity historical distribution in Section \ref{sec:FG}. The entities involved in ${Prompt}_{\text {replay }}$ are represented as a set $V_{\text {replay }}$, these entities are directly or indirectly influenced by newly arrived data:

\begin{equation}
\vspace{-0.3em}
    \resizebox{0.70\linewidth}{!}{
    \begin{math}
      \begin{aligned}
      V_{\text {replay }}=\left\{e \mid \forall(e, r, o) \in { Prompt }_{\text {replay }}\right\}\\ \cup\left\{e \mid \forall(s, r, e) \in  { Prompt }_{\text {replay }}\right\},
      \end{aligned}
    \end{math}
    }
  \label{eq:1}
\end{equation}

\subsection{Diffusion-enhanced Historical Distribution Generation}
\label{sec:FG}

The target of Diff-HDG strategy is to generate the historical distribution of entity with minimal conflicts against the current distribution of entity. For this purpose, during the generation process, common features between the historical and current distributions need to be enhanced.  
In addition, features in the historical distribution of entity that differ from the current distribution of entity need to be weakened.
Motivated by previous work~\citep{Reco,Sketch,Adding}, pre-trained DMs have demonstrated exceptional capabilities in reproducing knowledge from prompt texts. DMs possess a robust ability to generate generalized expressions. This capability is crucial for resolving conflicts that arise between different distributions. Thus, we generate historical distribution representations of entities through a pre-trained DM based on HCP. 
The generation of entity historical distribution primarily relies on the inverse diffusion process, which can be outlined as follows:
\begin{equation}
  \resizebox{0.7 \linewidth}{!}{
    \begin{math}
      \begin{aligned}
      H_{e}^{\text {replay }}=p_{\phi}\left(X_{n}, f_{\theta_{t}},  {Prompt}_{{replay}}\right),
      \end{aligned}
    \end{math}
    }
  \label{eq:1}
\end{equation}
where $X_{n}$ denotes the object to denoising, which is processed iteratively to yield the historical distribution of entity $H_{e}^{\text {replay }}$. The function $f_{\theta_{t}}$ represents the parameters of the TKGR model at the current time $t$.


In detail, for a fact \scalebox{0.9}{$(s,r,e_{q})\in {Prompt}_{replay}$}, we treat the entity $s$ and the relation $r$ as generation conditions. This condition-based generation method integrates information from historical neighbors and relations, enabling a more precise modeling of the historical distribution of entity:
\begin{equation}
    \resizebox{0.79\linewidth}{!}{
    \begin{math}
      \begin{aligned}
X_{n}=\operatorname{Condition}\left(S_{0}, R_{0}, Z\right), Z \sim \mathcal{N}(\mathbf{0}, I),
      \end{aligned}
    \end{math}
    }
  \label{eq:1}
\end{equation}
where \scalebox{0.9}{$S_{0}=\operatorname{Embedding}\left(s\right)$}, \scalebox{0.9}{$R_{0}=\operatorname{Embedding}\left(r\right)$}, and \scalebox{0.9}{$\operatorname{Condition}\left(\cdot\right)$} represents concatenation. The condition in $X_{n}$ can guide the DM to generate distributions that reflect the historical semantics of entities.

To enhance the common features between historical and current distribution representations, we propose a novel method to guide the generation process of historical entity representations:
\begin{equation}
    \resizebox{0.33\linewidth}{!}{
    \begin{math}
      \begin{aligned}
      X_{n-1}=p_{\phi}\left(X_{n}\right),
      \end{aligned}
    \end{math}
    }
  \label{eq:1}
\end{equation}
\begin{equation}
    \resizebox{0.53\linewidth}{!}{
    \begin{math}
      \begin{aligned}
      X_{n-1}=X_{n-1}+\gamma \frac{\partial \sigma}{\partial X_{n-1}},
      \end{aligned}
    \end{math}
    }
  \label{eq:1}
\end{equation}
\begin{equation}
    \resizebox{0.80\linewidth}{!}{
    \begin{math}
      \begin{aligned}
      \frac{\partial \sigma}{\partial X_{n-1}}=\nabla_{X_{n-1}} \sigma\left(f_{\theta_{t}}\left(X_{n-1},\left(s, r, e_{q}\right)\right)\right),
      \end{aligned}
    \end{math}
    }
  \label{eq:1}
\end{equation}
\noindent where $\sigma$ denotes the softmax function, $\gamma$ is a hyperparameter, and $X_{n-1}$ is the result of the first denoising step performed on $X_{n}$. This process produces a cleaner representation of $X_{n}$. After acquiring $X_{n-1}$, we evaluate the scores of historical facts in ${Prompt}_{replay}$ with the current TKGR model $f_{\theta_{t}}$. The gradient of scores is applied to optimize the generated historical distribution, ensuring that the scores of these historical facts are maximized at the current time.
Based on our empirical observations, adjacent timestamps in TKGs show only minor distribution differences. Since the model parameters $\theta_{t}$ for the current time can only be obtained after being updated at the current time, we approximate $\theta_{t}$ using $\theta_{t-1}$.

After $n$ iterations of denoising with $p_{\phi}$, we obtain the generated representation $X_{0}^{e_q}$ for the query entity. Similarly, the historical neighboring entity $s$ receives an updated representation $X_{0}^{s}$, influenced by the query entity $X_{0}^{e_q}$. 
Mean pooling is used to aggregate information from multiple neighbors across different timestamps, as shown below:
\begin{equation}
    \resizebox{0.54\linewidth}{!}{
    \begin{math}
      \begin{aligned}
      H_{e}^{\text {replay }}=\frac{\sum_{i=1}^{k} \sum_{\epsilon \in M_{e}^{i}} H_{\epsilon}^{i}}{\sum_{i=1}^{k}\left|M_{e}^{i}\right|},
      \end{aligned}
    \end{math}
    }
  \label{eq:1}
\end{equation}
where $H_{e}^{\text {replay }}$ represents the final historical distribution representation of entity $e \in V_{\text {replay }}$, capturing its historical characteristics in the TKGs. $H_{\epsilon}^{i}$ represents the entity representation $X_{0}^{e}$ generated from the facts $\epsilon$. $M_{e}^i$ refers to the set of facts that contain entity $e$ at the $i$-th time slice.
After iterative denoising, the features in $H_e^{replay}$ that are the same as the current distribution are enhanced, and the features in $H_e^{replay}$ that are different from the current distribution are weakened.
These historical entity representations are generated in parallel to improve computational efficiency.
\subsection{Deep Adaptive Replay}
\label{sec:DFR}
In this section, we introduce a DAR mechanism that effectively integrates the historical and current distributions of entities. Building upon the historical distribution representation of entities obtained in section \ref{sec:FG}, these representations are incorporated into the current distribution representation of entities. 

We identify that overly complex historical knowledge injection mechanisms impose a considerable learning burden, whereas excessively simplistic approaches result in significant knowledge loss. To overcome these issues, we propose DAR for historical knowledge replay, which performs the following operations at each layer of the KG snapshot sequence reasoning model:
\begin{equation}
    \resizebox{0.85\linewidth}{!}{
    \begin{math}
      \begin{aligned}
       H_{e}^{l}=\left\{\begin{array}{c}
 H_{e}^{\text {current}, l}, \quad e \notin V_{\text {replay }} \\
\alpha H_{e}^{\text {replay }}+ (1-\alpha)H_{e}^{\text {current}, l}, e \in V_{\text {replay }}
\end{array}\right.,
      \end{aligned}
    \end{math}
    }
  \label{eq:12}
\end{equation}
where $\alpha \in [0,1]$, which adaptively balances new and old knowledge. $H_{e}^{\text {current}, l}$ denotes the entity distribution representation of the current task at layer $l$. To preserve the evolutionary characteristics in the temporal sequence, deep replay is conducted within the $L$ evolution units; the final entity representation, denoted as $H_{\text {final }}$, is obtained.
\vspace{-0.5em}
\subsection{Model Training}
\label{sec:Optimization}
\begin{table*}[htbp]
  \centering
  \resizebox{\linewidth}{!}{
    \begin{tabular}{c|c|ccc|c|ccc|c|ccc}
    \toprule
    \multicolumn{1}{r}{} & \multicolumn{1}{r}{} & \multicolumn{3}{c}{ICE14} & \multicolumn{1}{c}{} & \multicolumn{3}{c}{ICE18} & \multicolumn{1}{c}{} & \multicolumn{3}{c}{ICE05-15} \\
    \midrule
    \multirow{2}[4]{*}{Algo.} & Current & \multicolumn{3}{c|}{Average} & Current & \multicolumn{3}{c|}{Average} & Current & \multicolumn{3}{c}{Average} \\
\cmidrule{2-13}          & MRR   & MRR   & Hits@1 & Hits@10 & MRR   & MRR   & Hits@1 & Hits@10 & MRR   & MRR   & Hits@1 & Hits@10 \\
    \midrule
    FT    & 42.66 & 37.46 & 26.95 & 58.20 & 30.76 & 25.35 & 15.97 & 44.36 & 43.80 & 41.88 & 30.55 & 63.82 \\
    ER    & 48.75 & 42.14 & 31.03 & 63.80 & 30.39 & 27.20 & 16.88 & 48.19 & 52.50 & 45.55 & 33.34 & 69.07 \\
    TIE   & 53.74 & 41.07 & 30.28 & 62.39 & 34.45 & 28.73 & 18.40 & 49.60 & 60.77 & 42.56 & 30.90 & 64.67 \\
    LKGE  & 43.56 & 37.51 & 27.13 & 58.51 & 31.12 & 25.56 & 16.12 & 44.70 & 43.28 & 42.46 & 30.99 & 64.51 \\
    IncDE & 45.03 & 36.57 & 26.20 & 56.95 & 31.83 & 25.52 & 16.07 & 44.74 & 46.33 & 40.56 & 29.34 & 62.17 \\
    \midrule
    \textbf{DGAR} & \textbf{58.59} & \textbf{50.12} & \textbf{39.36} & \textbf{70.48} & \textbf{36.53} & \textbf{33.00} & \textbf{21.74} & \textbf{55.63} & \textbf{66.01} & \textbf{54.33} & \textbf{43.11} & \textbf{75.13} \\
    \bottomrule
    \end{tabular}%
    }
  \caption{The main experimental results on the ICE14, ICE18, and ICE05-15 datasets are presented. Bolded scores indicate the best results.}
  \label{tab:mainResult1}%
  \vspace{-1.2em}
\end{table*}%
\begin{table}[htbp]
  \centering
  \resizebox{0.85\linewidth}{!}{
    \begin{tabular}{c|c|ccc}
    \toprule
    \multicolumn{1}{r}{} & \multicolumn{1}{c}{} & \multicolumn{3}{c}{GDELT} \\
    \midrule
    \multirow{2}[4]{*}{Algo.} & Current & \multicolumn{3}{c}{Average} \\
\cmidrule{2-5}          & MRR   & MRR   & Hits@1 & Hits@10 \\
    \midrule
    FT    & 14.74 & 15.60 & 8.73  & 29.05 \\
    ER    & 15.42 & 16.21 & 8.97  & 30.42 \\
    TIE   & 15.56 & 16.40 & 8.94  & 30.98 \\
    LKGE  & 14.43 & 15.52 & 8.69  & 28.90 \\
    IncDE & 15.14 & 15.49 & 8.64  & 28.86 \\
    \midrule
    \textbf{DGAR} & \textbf{23.25} & \textbf{28.30} & \textbf{17.38} & \textbf{51.39} \\
    \bottomrule
    \end{tabular}%
    }
  \caption{The main experimental results on the GDELT.}
  \label{tab:mainResult2}%
  \vspace{-0.5em}
\end{table}%

After obtaining the final representation, the decoder computes scores for candidate entities. We treat entity prediction as multi-class classification, and model parameters $\theta_{t}$ for task $t$ are optimized as follows:
\begin{equation}
    \resizebox{0.70\linewidth}{!}{
    \begin{math}
      \begin{aligned}
      \mathcal{L}_{t, c}=-\sum_{(s, r, o, t) \in D_{\text {train }}^{t}} y_{t}^{e} f_{\theta_{t}}(s, r, o, t),
      \end{aligned}
    \end{math}
    }
  \label{eq:1}
\end{equation}

\noindent where $y_{t}^{e}$ represents the label vector.
During experiments, we observe that although we attempt to preserve historical knowledge by enhancing common features between the representations of current and historical distributions, the model still suffers from historical information loss. This is primarily due to the constraints of the guidance function and subsequent optimization for current data. To address this issue, we incorporate facts from the historical context prompt as a regularization term into the loss function. The final loss calculation is formulated as follows:
\begin{equation}
    \resizebox{0.40\linewidth}{!}{
    \begin{math}
      \begin{aligned}
      \mathcal{L}_{t}=\mathcal{L}_{t, c}+\mu \mathcal{L}_{t, r},
      \end{aligned}
    \end{math}
    }
  \label{eq:1}
\end{equation}
where $\mathcal{L}_{t, c}$ represents the training loss for the current task, $\mathcal{L}_{t, r}$ denotes the loss associated with replaying historical facts, and $\mu$ is a hyperparameter, typically set to 1. The computation of $L_{t,r}$ is similar to that of $L_{t,c}$. The difference is that $L_{t,c}$ calculates the loss based on current facts, while $L_{t,r}$ uses historical fact in $Prompt_{replay}$.
\vspace{-0.2em}
\section{Experiments}
\vspace{-0.2em}
\subsection{Experimental Setup}
\textbf{Datasets.}\indent We adopt four widely used benchmark datasets for TKGR tasks: ICE14, ICE18, ICE05-15, and GDELT. The first three datasets originate from the Integrated Crisis Early Warning System~\citep{jin-etal-2020-recurrent}, which records geopolitical events. The statistical details of these datasets are summarized in Table \ref{sec:Datasets Details}. 

\noindent\textbf{Metrics.}\indent We utilize two evaluation metrics: Mean Reciprocal Rank (MRR) and Hits@k (k=1,10), both of which are widely adopted to assess the performance of TKGR methods. Following the approach~\citep{HistoryRepeat}, we evaluate the model's ability to mitigate catastrophic forgetting. We evaluate the model trained on the final task $t$ by testing its performance on the current test set (Current) and calculating its average performance across all previous test sets (Average).


\noindent\textbf{Baselines.}\indent We adopt the following baselines: FT, ER~\citep{rolnick2019experience}, TIE~\citep{wu2021tie}, LKGE~\citep{cui2023lifelong} and IncDE~\citep{IncDE}. Details about these baselines are provided in Appendix \ref{sec:Baselines Details}. In the experiments, we use RE-GCN as the base model.
\subsection{Main Results}
The results of main experiments are shown in Table \ref{tab:mainResult1} and Table \ref{tab:mainResult2}. Each dataset is tested five times and the average results are reported. The same procedure is also followed in subsequent experiments.

DGAR achieves consistent performance improvements compared to Fine-tuning.
For historical tasks, it achieves an average increase of 11.34\% in MRR. This demonstrates that, compared to direct fine-tuning, our approach more effectively retains historical knowledge.

Moreover, DGAR consistently outperforms all baselines. Compared to the strongest baseline, it achieves an average MRR improvement of 4.01\% in the current task across all evaluated datasets. For historical tasks, the average improvement is 8.23\% in MRR, and 9.79\% in Hits@10. DGAR demonstrates improvements across various datasets by preserving historical knowledge.

In contrast, while the TIE model performs well on current tasks, it exhibits poor average performance across all historical tasks. This result is attributed to the strict regularization of TIE, which limits its ability to retain historical knowledge. IncDE and LKGE only use the embedding constraint model of entities and relations at the previous moment to retain old knowledge, which leads to the decline of IncDE's performance on historical tasks compared to FT. LKGE additionally considers the constraints of cumulative weights, so it has a slight improvement on certain datasets compared to FT.
\begin{figure}[t]
    \centering
    \begin{subfigure}{0.49\linewidth}
        \centering
        \includegraphics[width=\linewidth]{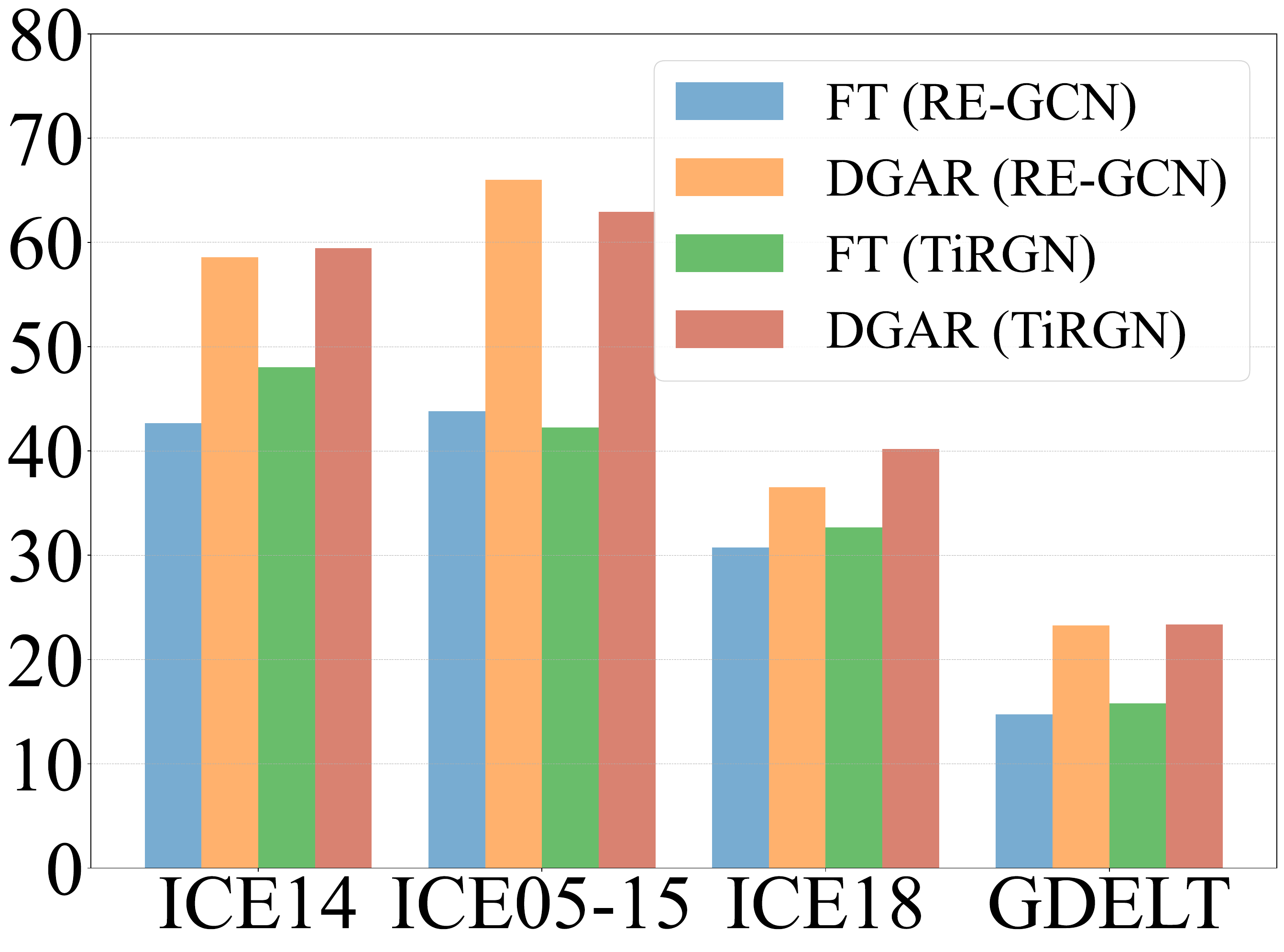}
        \caption{Current}
        \label{chutian1}
    \end{subfigure}
    \begin{subfigure}{0.49\linewidth}
        \centering
        \includegraphics[width=\linewidth]{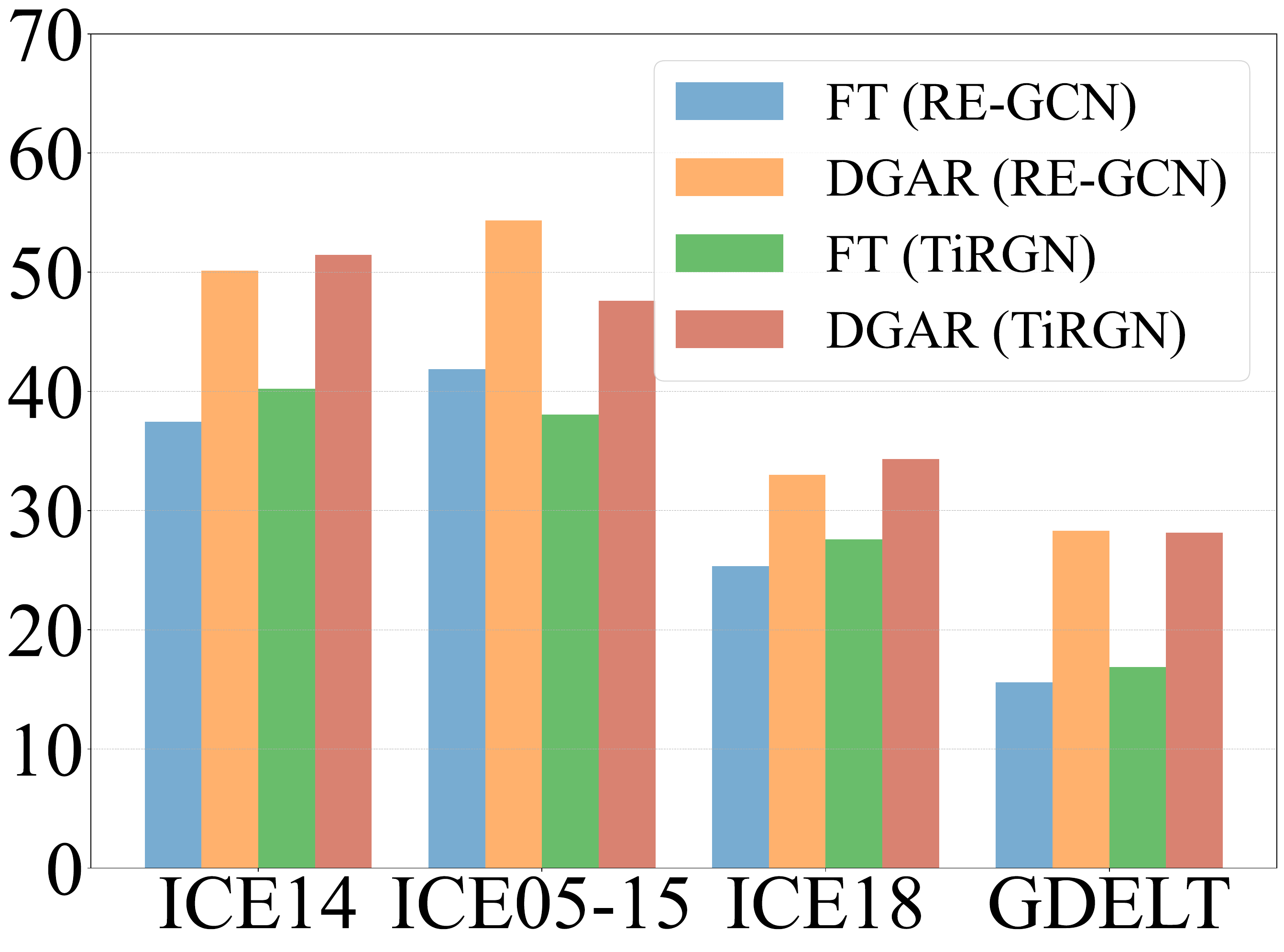}
        \caption{Average}
        \label{chutian2}
    \end{subfigure}
    \caption{Performance of different base TKGR models.}  
    \label{fig:test_TiRGN}
    \vspace{- 0.8em}
\end{figure}
\subsection{Ablation Study}
\begin{table*}[htbp]
  \centering
  \resizebox{\linewidth}{!}{
        \begin{tabular}{c|c|cc|c|cc|c|cc|c|cc}
    \toprule
    \multicolumn{1}{r}{} & \multicolumn{3}{c}{ICE14} & \multicolumn{3}{c}{ICE18} & \multicolumn{3}{c}{ICE05-15} & \multicolumn{3}{c}{GDELT} \\
    \midrule
    \multirow{2}[4]{*}{} & Current & \multicolumn{2}{c|}{Average} & Current & \multicolumn{2}{c|}{Average} & Current & \multicolumn{2}{c|}{Average} & Current & \multicolumn{2}{c}{Average} \\
\cmidrule{2-13}          & MRR   & MRR   & Hits@10 & MRR   & MRR   & Hits@10 & MRR   & MRR   & Hits@10 & MRR   & MRR   & Hits@10 \\
    \midrule
    w/o HP & 53.43 & 46.74 & 67.58 & 31.67 & 25.89 & 45.31 & 53.53 & 45.71 & 68.18 & 15.49 & 17.18 & 32.73 \\
    w/o GR &  48.18 & 39.15 & 59.71 & 30.32 & 27.62 & 47.74 & 52.97 & 38.71 & 59.19 & 16.24 & 16.67 & 31.65 \\
    w/o AR & 49.27 & 45.55 & 65.27 & 32.28 & 29.79 & 50.90 & 58.48 & 51.93 & 72.57 & 22.04 & 26.98 & 49.76 \\
    w/o Guider & 55.23 & 49.32 & 70.29 & 35.16 & 32.25 & 54.67 & 58.70 & 52.53 & 72.85 & 22.18 & 27.99 & 50.93 \\
    w/o $\mathcal{L}_r$ & 52.17 & 44.43 & 64.13 & 36.20 & 30.62 & 52.98 & 58.64 & 51.98 & 72.01 & 22.84 & 25.95 & 48.75 \\
    \midrule
    \textbf{Ours} & \textbf{58.59} & \textbf{50.12} & \textbf{70.48} & \textbf{36.53} & \textbf{33.00} & \textbf{55.63} & \textbf{66.01} & \textbf{54.33} & \textbf{75.13} & \textbf{23.25} & \textbf{28.30} & \textbf{51.39} \\
    \bottomrule
    \end{tabular}%
    }
    \caption{Ablation experimental results on all datasets.}
  \label{tab:ablation1}%
  \vspace{-1.0em}
\end{table*}%
In this section, we examine the impact of various components of the model on the final result, as shown in Table \ref{tab:ablation1}.
To thoroughly evaluate their roles, we implement the following model variants: (1) w/o HP, where the HCP is replaced with ER; (2) w/o GR, where the variant discards DAR and Diff-HDG, relying only on the facts within the HCP for regularization; (3) w/o AR, where the historical and current entity distributions are merged through direct addition instead of DAR as specified in Eq. \ref{eq:12}; and (4) w/o Guider, where the operation of enhancing common features across different distributions in Diff-HDG is discarded; (5) w/o $L_r$, where the loss $\mathcal{L}_{t,r}$ in Section \ref{sec:Optimization} is removed during training. The analysis of w/o $L_r$ in Appendix \ref{sec:L_r}.

\noindent\textbf{Effect Analysis of Historical Context Prompt.}\indent In the w/o HP variant, the model's performance noticeably declined, demonstrating that HCP effectively ensures the semantic integrity of the historical information. It prevents catastrophic forgetting during CL and enhances predictions for the current task. In contrast, ER merely replays partial historical information.

\noindent\textbf{Effect Analysis of Diffusion-enhanced Historical Distribution Generation.}\indent In the w/o Guider variant, different datasets show varying degrees of performance drop. This demonstrates that incorporating the guider aids in capturing common features between historical and current distributions, thereby mitigating performance losses caused by distribution conflicts. The smaller drop observed on the GDELT dataset likely results from its shorter temporal gaps and less pronounced distribution shifts compared to other datasets.

\noindent\textbf{Effect Analysis of Deep Adaptive Replay.}\indent Removing the adaptive parameter $\alpha$ in the w/o AR variant causes varying levels of performance decrease across datasets, demonstrating the effectiveness of our adaptive fusion method in balancing historical and current distribution representations. 
The limited decrease observed on the GDELT dataset can be attributed to the minimal difference between the current distribution and that of GDELT, which restricts the adaptive parameter $\alpha$ in its capacity to adjust effectively.

\noindent\textbf{Combined Effect Analysis of DAR and Dif-HDG.}\indent DAR and Diff-HDG are proposed to resolve the conflict between historical and current distribution.
The w/o GR variants show a clear performance drop, likely due to memory contention caused by distribution conflicts arising from the simple replay of historical data.
This suggests that Dif-HDG and DAR alleviate distribution conflicts, thereby preserving historical knowledge more efficiently.

\noindent\textbf{Effect Analysis of Different Base Models.}\indent Although we choose the most typical model, RE-GCN, as our base model, our method can still be extended to other GNN-based TKGR models. To verify the scalability of our method, we extend DGAR to TiRGN~\citep{TiRGN} and conduct experiments on four benchmark datasets. The experimental results of MRR in Figure \ref{fig:test_TiRGN} indicate that DGAR consistently outperforms direct fine-tuning on both current tasks and historical tasks, demonstrating that DGAR has robust scalability and effectiveness for GNN-based TKGR models.
\subsection{Effect of Memorizing in CL}
\begin{figure}[t]
    \centering
    \begin{subfigure}{0.48\linewidth}
        \centering
        \includegraphics[width=\linewidth]{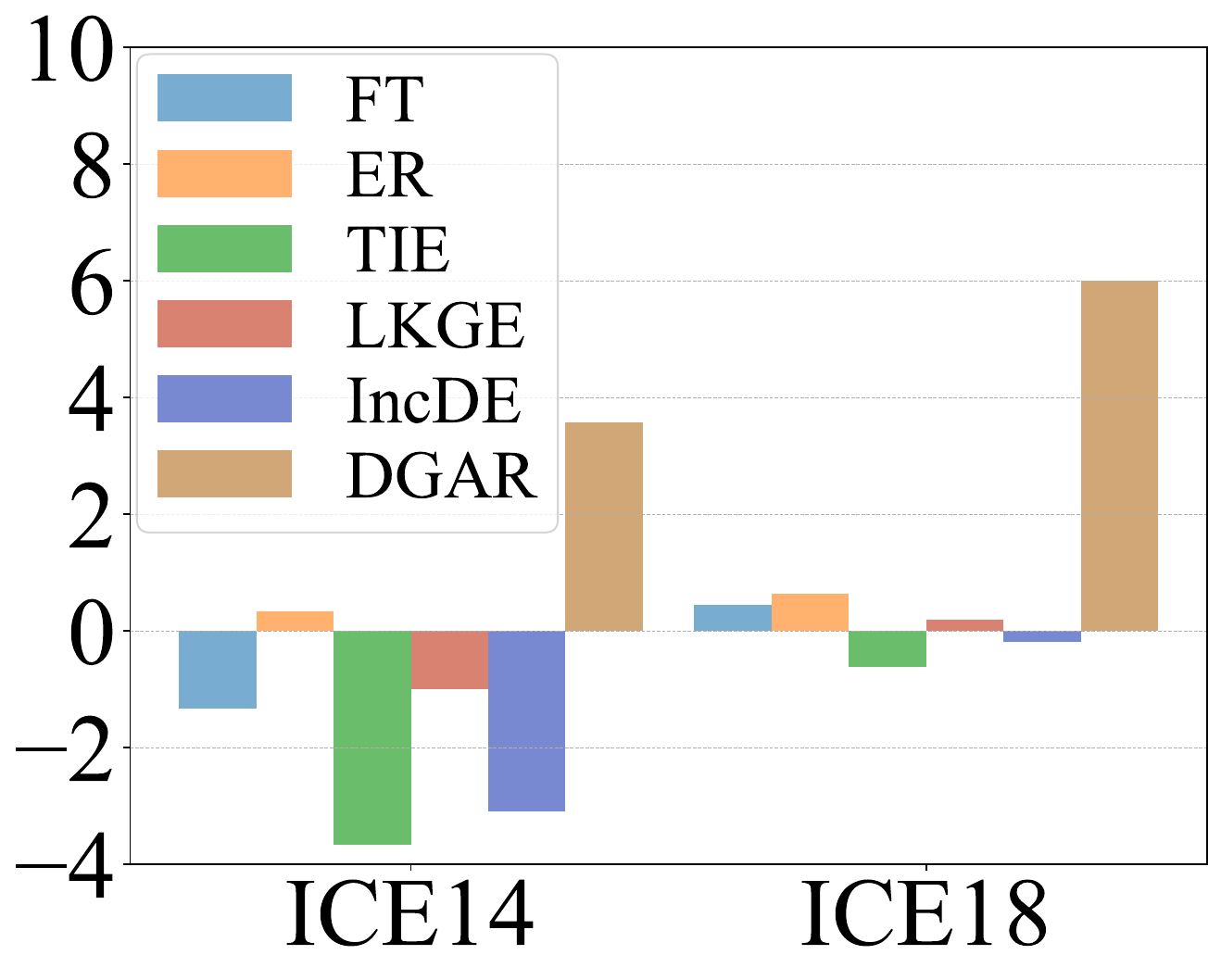}
        \caption{ICE14 and ICE18}
        \label{chutian1}
    \end{subfigure}
    \begin{subfigure}{0.49\linewidth}
        \centering
        \includegraphics[width=\linewidth]{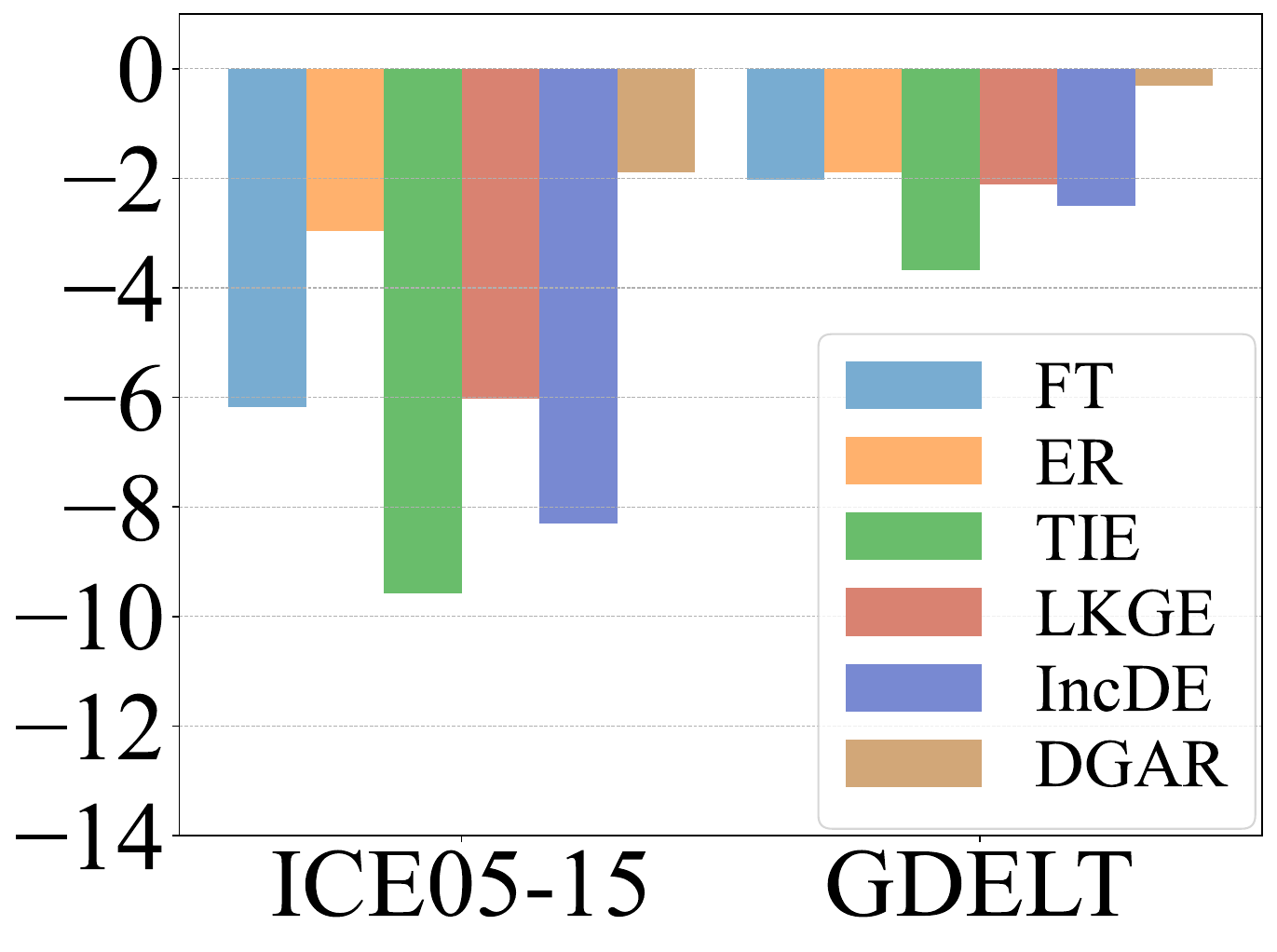}
        \caption{ICE05-15 and GDELT}
        \label{chutian2}
    \end{subfigure}
    \caption{Effect of memorizing old knowledge in CL.}  
    \label{fig:test_forget}
    \vspace{- 1.5em}
\end{figure}

To further validate DGAR's ability to retain historical knowledge during forward learning, we evaluate the mean difference between $p_{n,i}$ and $p_{i,i}$ ($1<i \leq n$) in Figure \ref{fig:test_forget}. Here, $p_{i,j}$ represents the MRR score of the $j$-th task after training the model on the $i$-th task. A higher mean value indicates better retention of prior knowledge during CL. When the value exceeds zero, it indicates reverse transfer of newly learned knowledge, whereas a value below zero reflects the loss of prior knowledge during CL~\citep{Lin2022BeyondNC}. Experiments show that DGAR outperforms the best baseline, confirming its effectiveness in mitigating catastrophic forgetting. 
Since the data in TKGs are highly correlated, we find that when the number of tasks is small, a reasonable strategy can help prevent catastrophic forgetting and facilitate the reverse transfer of new knowledge.
This is evident in the performance of DGAR and ER on ICE14.
\subsection{Case Study}
\begin{figure}[t]
    \centering
    \includegraphics[width=\columnwidth]{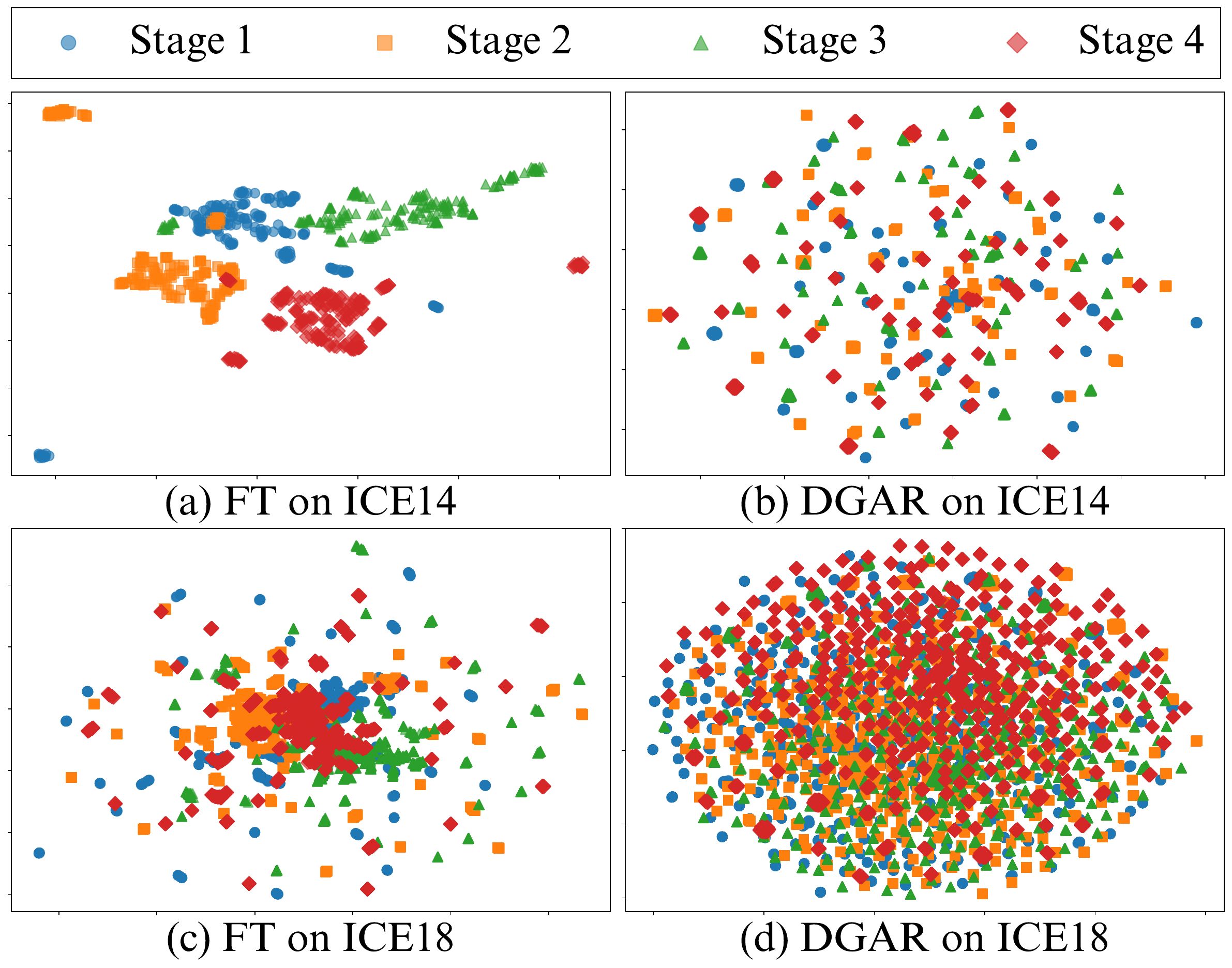}
    \caption{Visualization case study of entity distribution.}
    \label{fig:case study}
\end{figure}

We conduct a case study on ICE14 and ICE18 datasets to assess whether DGAR can handle conflicts in entity distributions and retain old knowledge effectively in Figure \ref{fig:case study}. At a randomly chosen time $i$, we extract all entities from the facts, save their feature distributions at time $i$ (Stage 1), time $j$ (Stage 2), time $m$ (Stage 3) and at a later time $n$ ($n>m>j>i$) (Stage 4), and analyzed them using U-MAP.

Figures \ref{fig:case study}(a) and \ref{fig:case study}(b) compare the entity distribution representations of the FT model and DGAR on ICE14 across four stages.
The entity distribution learned by the FT model at the same time is more clustered, while the entity distributions at different times are more distinct, showing a clearer difference.
In contrast, DGAR learns a more general and consistent distribution, allowing it to share the feature space between tasks more effectively, thereby enhancing knowledge retention and reducing forgetting.
A similar pattern is observed in Figures \ref{fig:case study}(c) and \ref{fig:case study}(d) on ICE18, further supporting these findings.
	
\vspace{-0.7em}
\section{Conclusion}
\vspace{-0.7em}
This paper introduces a deep generative adaptive replay method to mitigate catastrophic forgetting in TKGR models during CL.
A historical context prompt integrating contextual information is designed to generate historical distribution representations of entities via a pre-trained DM. 
The generation process is guided by current model parameters to reinforce common features, minimizing conflicts between historical and current entity distributions.
In addition, a deep adaptive replay strategy derives entity distribution representations with historical knowledge.
These combined techniques enable the proposed method to achieve outstanding performance across various datasets.

\section{Limitations}
In this section, we examine the limitations of our approach. DGAR is designed to retain previously acquired knowledge through CL, facilitating TKGR. Although DGAR is more time-efficient than retraining and surpasses other models in mitigating catastrophic forgetting, it still faces several challenges.

Firstly, the model addresses newly emerging entities and relations using Xavier initialization without further analysis or dedicated modeling. Such a simplistic approach may constrain the model's ability to learn new knowledge effectively, particularly when complex interrelations exist between new and previously learned knowledge. This highlights the need for more sophisticated strategies to handle new entities and relations in CL scenarios.

Secondly, while DGAR demonstrates strong performance in reducing catastrophic forgetting, it introduces additional learnable parameters. These parameters enhance adaptability to new knowledge but also pose a potential risk of forgetting previously learned information. This risk arises since the increased number of parameters may lead the model to prioritize new knowledge, thereby compromising the retention of older knowledge. Furthermore, the inclusion of additional parameters inherently increases model complexity, making the training and reasoning process more cumbersome. Such complexity necessitates careful consideration during design to strike a balance between knowledge retention and model complexity.

\section{Ethics Statement}
Firstly, this study fully complies with the ethical guidelines in the ACL Code of Ethics. Secondly, all datasets involved in this study are from previous studies. The datasets we used do not contain individual privacy data. Finally, DGAR focuses on the research and experiments of TKGR tasks. Like other TKGR methods, the results of our method reasoning may be toxic or erroneous, so manual inspection of the results may be required in the applications.

\bibliography{custom}

\appendix
\section{Details about DGAR}
\subsection{Details about Deep Adaptive Replay}
\label{sec:Details about Deep Adaptive replay}
While initially exploring more complex mechanisms for this integration, it is observed that the additional parameters introduced hinders the model’s convergence due to increased fitting complexity. As a result, a direct injection approach is adopted to integrate the historical distributions into the current representations, as detailed below:
\begin{equation}
    \resizebox{0.80\linewidth}{!}{
    \begin{math}
      \begin{aligned}
      H_{\text {final }}=\left\{\begin{array}{c}
H_{e}^{\text {current },}, \quad e \notin V_{\text {replay }} \\
H_{e}^{\text {replay }}+H_{e}^{\text {current }}, e \in V_{\text {replay }}
\end{array}\right..
      \end{aligned}
    \end{math}
    }
  \label{eq:12}
\end{equation}

However, it is observed that such a simple and direct fusion approach results in performance degradation. To address this issue, a straightforward parameter is introduced to balance the historical distribution representation $H_{e}^{\text {replay }}$ and the current distribution representation $H_{e}^{\text {current }}$, thereby generating the final entity representation $H_{\text {final }}$. This parameter effectively adjusts the weighting of the two distributions, mitigating the performance loss caused by direct fusion while preserving the expressive power of historical knowledge and the dynamic characteristics of the current distribution:
\begin{equation}
    \resizebox{0.85\linewidth}{!}{
    \begin{math}
      \begin{aligned}
      H_{\text {final }}=\alpha H_{e}^{\text {replay }}+(1-\alpha) H_{e}^{\text {current}}, e \in V_{\text {replay }},
      \end{aligned}
    \end{math}
    }
  \label{eq:16}
\end{equation}
where $\alpha \in [0,1]$. $H_{\text {final }}$ denotes the final entity representation, combining the most recent and historical information of the entity.

To prevent the loss of entity evolution patterns over time caused by direct injection, we integrate distribution representation from the KG snapshot sequence reasoning model's evolution unit to achieve a deeper incorporation of historical distribution representations without introducing additional parameters. After applying the relation-aware GCN in the $l$-th evolution unit, we obtain the current distribution representation of entity $H_{e}^{\text {current}, l}$. The historical distribution representation of entities is then injected into the current distribution representation of entities as follows:
\begin{equation}
    \resizebox{0.85\linewidth}{!}{
    \begin{math}
      \begin{aligned}
      H_{e}^{\text {current}, l}=\alpha H_{e}^{\text {replay }}+(1-\alpha) H_{e}^{\text {current}, l}, e \in V_{\text {replay }}.
      \end{aligned}
    \end{math}
    }
  \label{eq:1}
\end{equation}

After passing through multiple evolution units, the final entity representation $H_{\text {final }}$, which incorporates historical distributions, is obtained.

\subsection{Pre-train for DM}
In this section, we will discuss how we obtain the pre-trained DM. Considering that using a large amount of knowledge to train diffusion will not only increase the risk of data leakage, but also fail to adapt to new data arriving over time, we adopt CL to train DM. For example, at the $t$-th moment, we can obtain the DM $\phi_{t-1}$ pre-trained at the previous moment, and $\phi_{t-1}$ is used as a pre-trained DM to assist in generating the entity history distribution in Diff-HDG. After completing all the operations in Section \ref{sec:DGAR}, $\phi_{t}$ is initialized with $\phi_{t-1}$, and the model parameters are updated on $D_{train}^t$. Eq. \ref{eq:16} is employed to preserve the historical knowledge of the entity for DM. Upon completion of the training, a new pre-trained DM, $\phi_{t}$, is obtained and will be used in the subsequent learning process.

\section{Further Analysis}
\footnotetext[2]{AI such as GPT only assists us in translation and grammar checking.}
\label{sec:Experiments Detail}

\subsection{Datasets Details}
\label{sec:Datasets Details}
\begin{table}[htbp]
  \centering
  \resizebox{\linewidth}{!}{
    \begin{tabular}{ccccc}
    \toprule
          & \textbf{ICE14} & \textbf{ICE18} & \textbf{ICE05-15} & \textbf{GDELT} \\
    \midrule
    \textbf{Entities} & 6869  & 23033 & 10094 & 7,691 \\
    \textbf{Relations} & 230   & 256   & 251   & 240 \\
    \textbf{Tasks} & 365   & 304   & 4017  & 2,751 \\
    \textbf{Task granularity} & 24 hours & 24 hours & 24 hours & 15mins \\
    \textbf{Total number of train} & 74,845 & 373,018 & 368,868 & 1,734,399 \\
    \textbf{Total number of valid} & 8,514 & 45,995 & 46,302 & 238,765 \\
    \textbf{Total number of test} & 7,371 & 49,545 & 46,159 & 305,241 \\
    \bottomrule
    \end{tabular}%
    }
  \caption{Details of the TKG datasets.}
  \label{tab:datasetDetail}
\end{table}%
We follow the common division ratio of TKGR tasks: The facts in each task are partitioned into train, valid, and test sets in a ratio of 8:1:1~\citep{RE-GCN,CENET}. The statistical details of the datasets are shown in Table \ref{tab:datasetDetail}.

\subsection{Baselines Details}
\label{sec:Baselines Details}
FT (fine-tuning) is a naive baseline where the model is fine-tuned using newly added facts without any mechanism to alleviate catastrophic forgetting. 
FT is set up following the previous works such as TIE~\citep{wu2021tie}, LKGE~\citep{wu2021tie}, and IncDE~\citep{wu2021tie}.  FT enables the base model (e.g., RE-GCN and TiRGN) to perform CL without applying any additional strategies. Specifically, the base model inherits the parameters from the previous time step $i-1$ and continual training on the training data $D_{train}^t$ at time step $i$.
ER~\citep{rolnick2019experience} mitigates forgetting by replaying a subset of previously stored events alongside newly added facts during training. TIE~\citep{wu2021tie} incorporates temporal regularization, experience replay with positive facts, and the use of deleted facts as negative examples to effectively address both catastrophic forgetting and intransigence. LKGE~\citep{cui2023lifelong} preserves historical knowledge by leveraging historical weights and embeddings through the L2 paradigm. We incorporate the reconstruction loss and embedding regularization from LKGE into our objective function. When initializing new entities, their embedding transfer strategies are adopted. Based on the method proposed by IncDE~\citep{IncDE}, we leverage the hierarchical ordering measure of IncDE and incorporate the distillation loss proposed by IncDE into our objective function.

\subsection{Implementation Details}
\label{sec: Implementation Details}
For all datasets, the embedding size $d$ is set to 200, the learning rate $lr$ is set to 0.001, and the batch size is determined by the number of facts at each time step. The number of layers of the Transformer encoder for all datasets is set to 2. The temperature coefficient $\tau$ for all datasets is set to 0.5. The parameters of DAGR are optimized by using Adam during the training process. The optimal coefficient $\gamma$ in the Diff-HDG is set to 1. The optimal number of layers $L$ in the DAR is set to 3. The optimal loss coefficient $\mu$ in model training is set to 1. The number of samples of the best HCP $k$ for ICE14, ICE18, ICE05-15, and GDELT is set to 35, 25, 40, and 32, respectively. We conduct hyperparameter search experiments on the primary parameters of DGAR using control variables. The number of parameters in ICE14, ICE18, ICE05-15, and GDELT is 17.55 MB, 33.71 MB, 21.44 MB, and 21.38 MB, respectively.
All experiments are conducted on NVIDIA A40.
\subsection{Sensitivity Analysis}
\label{sec: hype}
\begin{figure}[t]
    \centering
    \begin{subfigure}{0.49\linewidth}
        \centering
        \includegraphics[width=\linewidth]{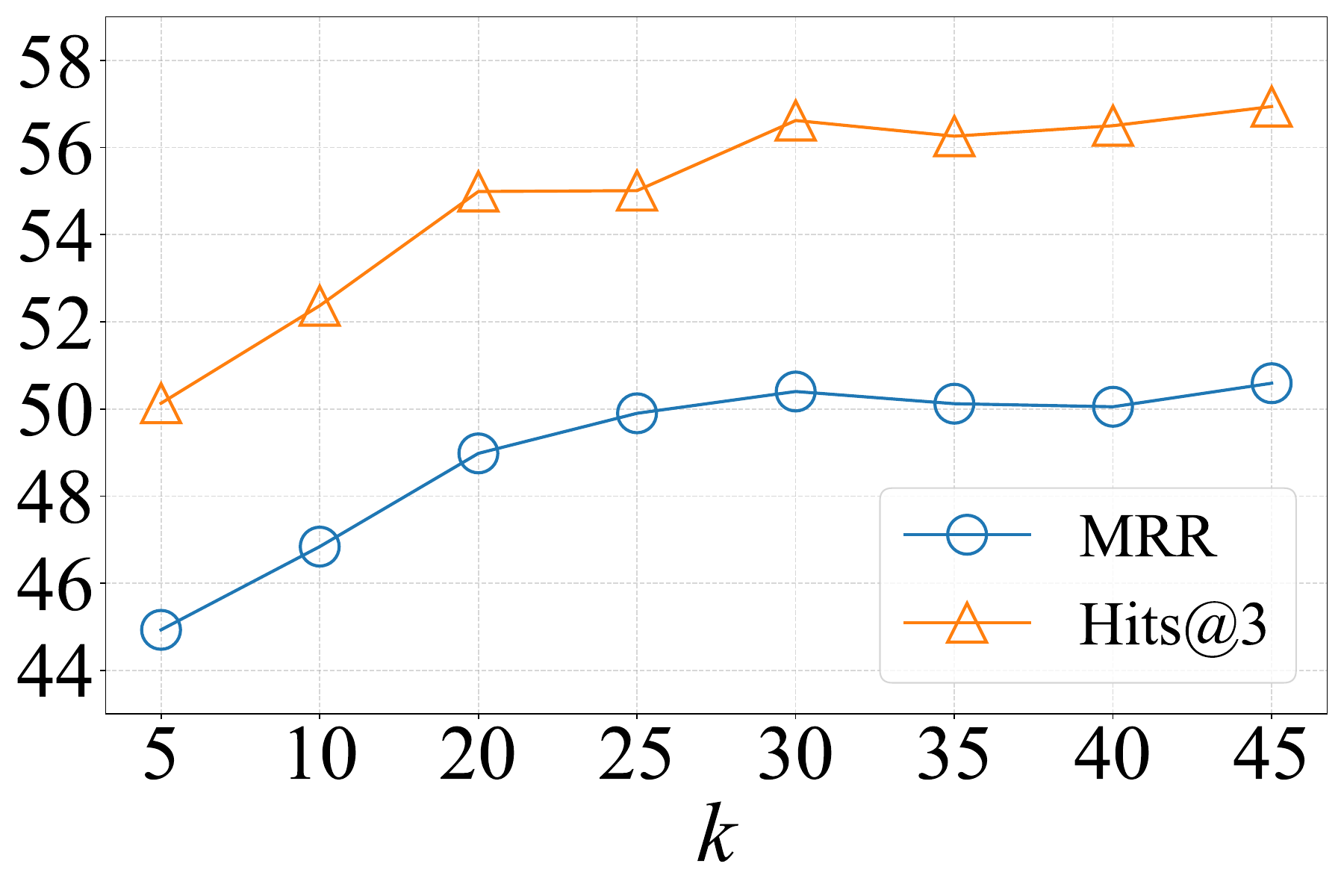}
        \caption{ICE14}
        \label{chutian1}
    \end{subfigure}
    \begin{subfigure}{0.49\linewidth}
        \centering
        \includegraphics[width=\linewidth]{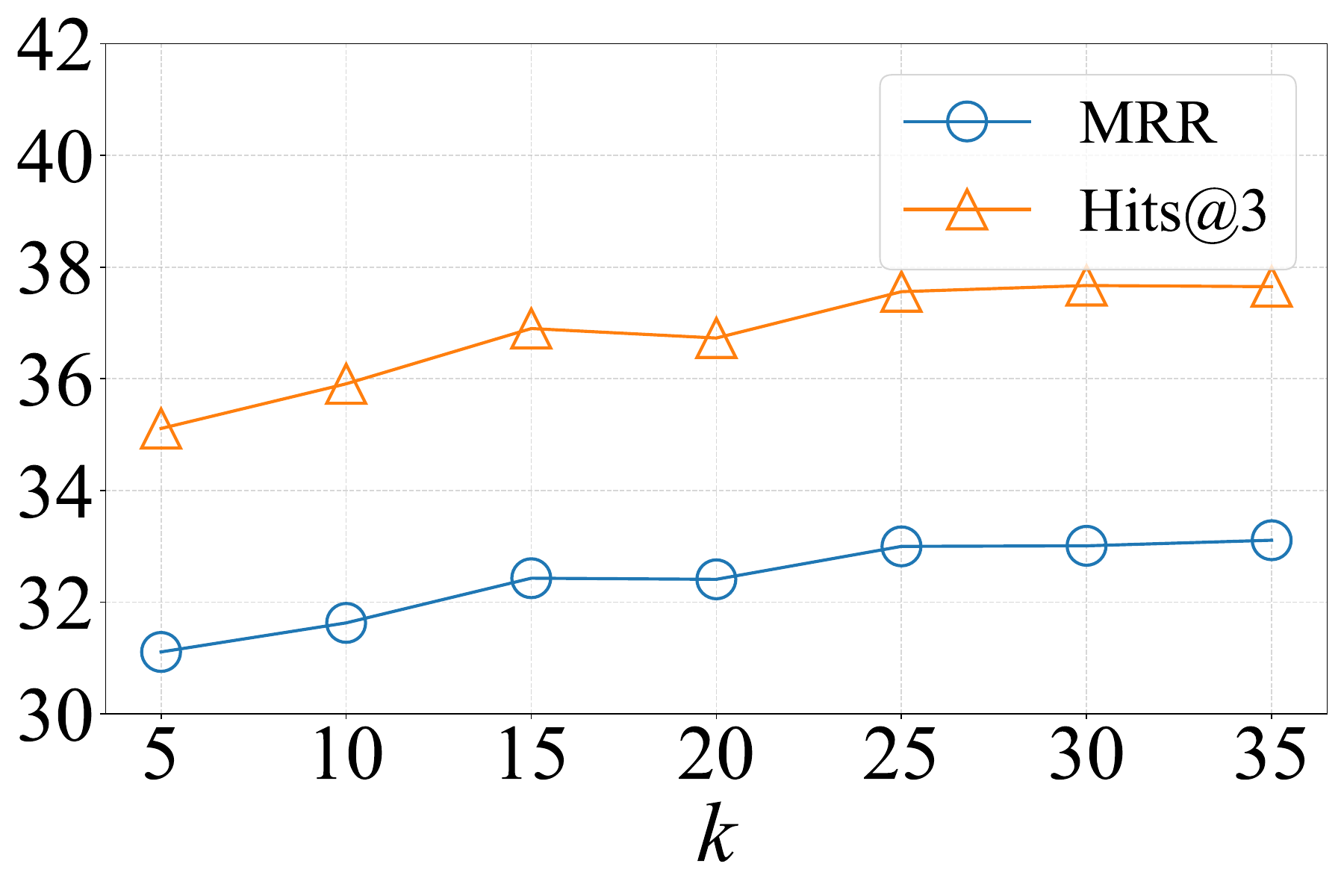}
        \caption{ICE18}
        \label{chutian2}
    \end{subfigure}
    \caption{Sensitivity Analysis.}  
    \label{fig:hype}
\end{figure}

In this section, we conduct experiments on ICE14 and ICE18 to further analyze the impact of hyperparameter $k$ in DGAR. The hyperparameter $k$ means the replay data consists of HCPs at $k$ different times.

To explore how the number of $k$ affects the model's ability to retain historical knowledge, we set different values for $k$. The results are shown in Figure \ref{fig:hype}, including MRR and Hits@3 results on the historical tasks. A larger k means the replay data consists of HCPs at more different times.
The results reveal that on the ICE14 dataset, DGAR demonstrates an initial improvement followed by a plateau as $k$ increases. These results indicate that selecting HCPs at more different times for historical distribution representations replay does not significantly enhance the final performance but instead increases computational costs. Notably, even with only 5 recall time slices, DGAR outperforms all baseline models. This demonstrates that our approach can effectively help the model retain historical knowledge, even under limited memory constraints.
\subsection{Inference Efficiency}
\label{sec:Inference Efficiency}
\begin{table}[htbp]
  \centering
  \resizebox{\linewidth}{!}{
    \begin{tabular}{c|ccc|ccc}
    \toprule
    \multicolumn{1}{r}{} & \multicolumn{3}{c}{ICE14} & \multicolumn{3}{c}{ICE18} \\
    \midrule
          & $k$     & Time(s) / Task & MRR   & $k$     & \multicolumn{1}{l}{Time(s) / Task} & MRR \\
    \midrule
    Retrain & \textbf{—} & 553.48 & 49.80 & —     & 530.48 & 31.35 \\
    \midrule
    \multirow{4}[2]{*}{DGAR} & 5     & 4.00  & 44.93 & 5     & 13.12 & 31.11 \\
          & 25    & 5.21  & 49.90 & 15    & 16.32 & 32.43 \\
          & 35    & 5.42  & 50.12 & 25    & 18.83 & 33.00 \\
          & 45    & 5.68  & 50.05 & 35    & 19.41 & 33.11 \\
    \midrule
    FT    & —     & 2.53  & 37.46 & —     & 5.48  & 25.35 \\
    ER    & —     & 2.86  & 42.14 & —     & 5.16  & 27.20 \\
    TIE   & —     & 4.38  & 41.07 & —     & 10.71 & 29.31 \\
    LKGE  & —     & 2.70  & 37.51 & —     & 5.03  & 25.56 \\
    IncDE & —     & 4.07  & 36.57 & —     & 9.71  & 25.52 \\
    \bottomrule
    \end{tabular}%
    }
  \caption{Inference efficiency analysis.}
  \label{tab:Inference Efficiency}
\end{table}%
We report the average time cost of each task and the MRR on historical tasks for DGAR at different $k$ values in Table \ref{tab:Inference Efficiency}. The hyperparameter $k$ means the replay data consists of HCPs at $k$ different times. We further report the average time consumption of each task and the MRR on historical tasks under different baselines and retaining settings in Table \ref{tab:Inference Efficiency}. Unlike DGAR and the baselines, retraining involves reprocessing the entire dataset whenever new data arrives. By comparing the results, our method outperforms retraining and requires less time. This shows the powerful ability of our model in dealing with catastrophic forgetting. Because our model uses more complex operations than the baseline in order to retain more historical knowledge, it is more time-consuming. Future research will focus on enhancing reasoning efficiency while preserving the accuracy of historical knowledge retention.

\subsection{Effect Analysis of $\mathcal{L}_r$}
\label{sec:L_r}
In the w/o $\mathcal{L}_{r}$ variant, the performance of the historical tasks drops significantly. This shows that the $\mathcal{L}_{r,t}$ loss in Section \ref{sec:Optimization} effectively alleviates the loss of historical information caused by current data optimization and Diff-HDG.

\subsection{Random Selection of HCPs}
In order to verify whether the generalization of replay data is enhanced, we added an additional experiment to prove. Compared to replaying historical data from the specified $k$ time slices, randomly selecting across the entire history provides more global and generalizable information. Therefore, we specifically added an experiment where the historical context prompts from the $k$ nearest time slices was selected as the replay data. The experimental results are in the Table \ref{tab:Random Selection}~:
\begin{table}[htbp]
  \centering
  \resizebox{0.8\linewidth}{!}{
    \begin{tabular}{c|cc|c|cc}
    \toprule
    \multicolumn{3}{c}{ICE14} & \multicolumn{3}{c}{ICE18} \\
    \midrule
    k     & \multicolumn{2}{c|}{Average (MRR)} & k     & \multicolumn{2}{c}{Average (MRR)} \\
    \midrule
          & nearest	 & random &       & nearest	 & random \\
    \midrule
    25    & 46.64 & 49.90 & 15    & 31.23 & 32.43 \\
    35    & 48.11 & 50.12 & 25    & 31.89 & 33.00 \\
    45    & 47.86 & 50.05 & 35    & 32.03 & 33.11 \\
    \bottomrule
    \end{tabular}%
    }
  
  \caption{Effect of Random Selection}
  \label{tab:Random Selection}
\end{table}%

The nearest refers to replaying with historical context prompts sampled from the nearest $k$ time slices. As shown in the Table~\ref{tab:Random Selection}, regardless of different $k$ values, the random outperforms the nearest on the historical tasks. The experimental results are consistent with our expectations. Mainly because  the random selection of historical context prompts provides the model with more generalized data, thus improving the model's performance on the test set.

\subsection{Compare Base on LogCL}

To further verify whether DGAR can enhance the performance of recent GNN-based models under the CL setting, we conducted the following experiment.

We selected LogCL~\citep{Log_CL} , a representative GNN-based TKGR model from recent works, as the base model. Below, we report its performance under the CL setting (FT) and its performance when combined with DGAR in the same setting on two datasets.
\begin{table}[htbp]
  \centering
  
  \resizebox{\linewidth}{!}{
    \begin{tabular}{c|c|cc|c|cc}
    \toprule
    \multicolumn{1}{r}{} & \multicolumn{3}{c}{ICE14} & \multicolumn{3}{c}{ICE18} \\
    \midrule
    \multirow{2}[4]{*}{Algo.} & Current & \multicolumn{2}{c|}{Average} & Current & \multicolumn{2}{c}{Average} \\
\cmidrule{2-7}          & MRR   & MRR   & Hits@10 & MRR   & MRR   & Hits@10 \\
    \midrule
    FT    & 31.63 & 28.93 & 48.61 & 38.46 & 30.47 & 55.52 \\
    DGAR  & \textbf{37.12} & \textbf{33.08} & \textbf{53.83} & \textbf{43.20} & \textbf{32.88} & \textbf{58.88} \\
    \bottomrule
    \end{tabular}%
    }
  \label{tab:addlabel}%
  \caption{Performance base on LogCL}
\end{table}%

The above experiments show that DGAR significantly enhances LogCL’s reasoning performance under the CL setting. Interestingly, LogCL performs worse on the ICE14 dataset than on ICE18, which contrasts with its performance in the full-training setting. This discrepancy occurs because ICE14 has a simpler data distribution compared to ICE18, while LogCL’s complex model structure makes it prone to overfitting on ICE14. Under the CL setting, highly complex models struggle to maintain stable learned features as new data is introduced~\citep{lee2024impactmodelsizecatastrophic}. DGAR mitigates this issue by replaying historical information, helping LogCL retain its learned features more effectively. Consequently, TKGR methods that excel in full-training scenarios may not necessarily achieve better reasoning performance under CL setting.

\end{document}